\documentclass[a4paper]{tCPH2e}
\pdfoutput=1 
\usepackage[T1]{fontenc} 
\usepackage{amsfonts,amssymb,amsmath}
\usepackage{color}
\usepackage{subfigure}

\begin{document}

\title{Higgs boson cosmology}

\author{
\name{Ian G.\ Moss}
\affil{School of Mathematics and Statistics, Newcastle University, 
Newcastle Upon Tyne, NE1 7RU, U.K.}
}

\maketitle

\begin{abstract}
The discovery of the Standard Model Higgs boson opens
up a range of speculative cosmological scenarios, from the formation of structure 
in the early universe immediately after the big bang, to relics from the electroweak 
phase transition one nanosecond after the big bang, on to the end of the 
present-day universe through vacuum decay. Higgs physics is wide-ranging,
and gives an impetus to go beyond the Standard Models of particle physics and
cosmology to explore the physics of ultra-high energies and quantum gravity. 
\end{abstract}


\section*{Acknowledgements} 

This work was supported by the STFC under Consolidated Grant ST/J000426/1.

\section{Introduction}

The discovery of the Higgs boson \cite{Chatrchyan:2012ufa,Aad:2012tfa} has 
completed the list of fermions (quarks and leptons), vector bosons 
(photons, gluons, $W$ and $Z$) and the single scalar boson
that make up the Standard Model of particle physics \cite{quigg2007}. The Higgs has
given us the first glimpse of a particle associated with a fundamental scalar field. 
This has a special significance to cosmologists, because so many of our modern 
theories of the early universe depend on the unique features of
scalar field physics.

We know that universe was once incredibly hot. As the universe cooled, it most likely 
went through a sequence of phase transitions. The Standard Model Higgs field has a special
association with an electroweak phase transition, which took place at a temperature equivalent 
to an energy scale of $163\,{\rm GeV}$ around one nanosecond after the big bang.
If we are fortunate, phase transitions like this can leave behind signals or relics which give
us information about physics beyond the Standard Model.

There is a great deal of interest in a phase transition which happened far earlier, 
at a time around $10^{-35}\,{\rm s}$ after the big bang. According to the inflationary 
scenario, the universe was dominated then by the vacuum energy of a scalar field, causing it to 
undergo a period of exponential expansion \cite{guth81,linde82,albrecht82}. 
Originally, the idea was that this field would be the Higgs field in a Grand Unified Theory uniting the 
strong, weak and electromagnetic forces. It is easier for the exponential expansion to get going when
the energy scale of the phase transition is very high, and this fits in well with the idea of Grand Unification 
at around $10^{15}\,{\rm GeV}$.

Inflationary models lead to a natural explanation for the origin of the large scale structure
of the universe today. We measure this large scale structure primarily through observations
of cosmic microwave background (CMB) radiation and through galaxy surveys. The CMB gives 
us a snapshot of the universe from the time when it first became transparent,
around $370,000\,{\rm yr}$ after the big bang. Galaxy surveys (together with distortions of the
CMB) probe the large scale structure from just after the origin of the CMB up until the present day. 
In inflationary models, small quantum or thermal fluctuations from 
the era of inflation are frozen into the geometry of spacetime. These fluctuations are the cause of 
the intensity fluctuations in the cosmic microwave background. Different models of inflation,
with different dependence of the vacuum energy on the scalar field, give different predictions
for the cosmic microwave background fluctuations and allow us to test models of inflation.
  
The large vacuum energy needed for inflation can arise in models where large fields 
correspond to large vacuum energies, provided that the universe started out with a large-magnitude 
scalar field. This is achieved in the chaotic inflationary scenarios,
where the field starts out randomly distributed throughout space \cite{Linde1983177}. 
By chance, some small region has a large magnitude field and this region inflates. Eventually,
with enough inflation, this region grows to encompass our entire observable universe.
Once we get into the setting of these chaotic scenarios, then even a field normally associated
with energy scales well below $10^{15}\,{\rm GeV}$, like the Standard Model Higgs field, 
can drive inflation. This is the main topic covered in section \ref{sec3}.

Finally, we might enquire whether all the phase transitions are in the distant past, or whether we
live in a metastable state, or `false vacuum', which could decay into a new vacuum state \cite{Sher:1988mj}.
For the observed Higgs mass around $125\,{\rm GeV}$, the vacuum structure of the Higgs
field in the Standard Model becomes rather sensitive to couplings to other massive particles.
One possible configuration is a metastable state with the Higgs field having a value around 
$246\,{\rm GeV}$ and a true vacuum with the Higgs field close to the Planck scale $2.4\times 10^{18}{\rm GeV}$.
This would, of course, be very susceptible to the effects of new physics beyond the Standard Model.

Decay of our `false' vacuum state would be catastrophic, since it would change the masses and 
interactions of all the elementary particles. The fact that our vacuum has survived $13.8\,{\rm Gyr}$
since the big bang is evidence that the decay rate must be negligible. Vacuum decay is a quantum 
tunnelling process which is exponentially suppressed, so very long lifetimes for metastable
states are not unusual. On the other hand, everyday phase transitions are often triggered by
some kind of seed, a speck of ice or dust in a cloud for example causing water to condense
into droplets. In the last section we shall report on recent results which consider whether a tiny 
black hole could act as a nucleation seed and trigger the decay of our vacuum state. 

The theory of elementary particles is greatly simplified by using a special set
of units in which $\hbar=c=1$. The fundamental unit is then the 
${\rm GeV}=10^9\,{\rm eV}=0.1602\,{\rm J}\,(\rm{S.I.})$. In these units masses are given in GeV,
distances in ${\rm GeV}^{-1}=0.1973\,{\rm fm}\,({\rm S.I.})$.
The gravitational constant is replaced by the reduced Planck mass, 
$M_p=(8\pi G)^{-1/2}=2.435\times 10^{18}{\rm GeV}=4.3415\times 10^{-9}\,{\rm kg}\,({\rm S.I.})$.

\section{The Higgs field}\label{sec1}

\begin{figure}
\begin{center}
\resizebox*{14cm}{!}{\includegraphics{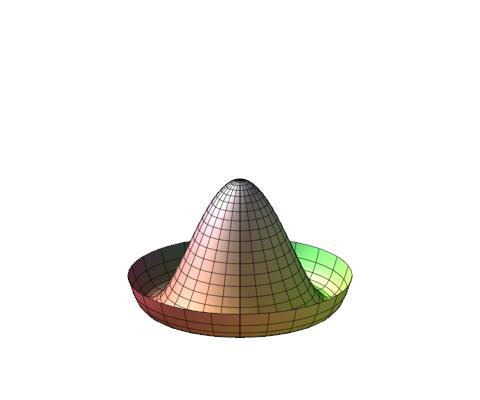}}
\caption{The famous `Mexican hat' shape for the Higgs potential. 
The potential is plotted vertically and solid lines indicate constant $(\phi,\alpha)$ 
parameters for the Higgs field, 
leaving two angles $\beta$ and $\gamma$ which are not shown. The Higgs
vacuum corresponds to a field lying somewhere along the brim of the hat.
The potential there has to be very close to zero, otherwise it would produce  a
large cosmological constant.} \label{higgs}
\end{center}
\end{figure}

We start with an overview of some of the essential features of Higgs field
theory in the part of the Standard Model which unites weak and electromagnetic forces:
the Weinberg-Salam model\cite{Langacker:2009my,0954-3899-37-7A-075021}. 
This model has an $SU(2)\times U(1)$ local symmetry group.

The unified Higgs field consists of two complex scalar fields arranged into a doublet
${\bf H}({\bf x},t)$. The Higgs field is a scalar under spatial rotations, but the components mix under
the $SU(2)\times U(1)$ symmetry group. The energy density of a stationary Higgs field 
configuration defines the Higgs potential $V({\bf H})$,
\begin{equation}
V({\bf H})=V_0-\mu^2{\bf H}^\dagger{\bf H}+\lambda({\bf H}^\dagger{\bf H})^2.
\end{equation}
Adding the constant $V_0$ has no effect on the particle physics, but it becomes important
for cosmology because all forms of energy affect the expansion of the universe.

In exactly the same way in which we might represent a vector by its magnitude and direction,
we can represent the Higgs field by a magnitude $\phi$ and rotation angles $\alpha$,
$\beta$ and $\gamma$,
\begin{equation}
{\bf H}={1\over\sqrt{2}}{\bf U}
\begin{pmatrix}
0\\
\phi \\
\end{pmatrix},
\qquad
{\bf U}=
\begin{pmatrix}
\cos\alpha+i\sin\alpha\cos\beta&ie^{-i\gamma}\sin\alpha\sin\beta\\
ie^{i\gamma}\sin\alpha\sin\beta&\cos\alpha-i\sin\alpha\cos\beta\\
\end{pmatrix}.
\end{equation}
The potential has the `Mexican hat' shape shown in figure \ref{higgs}, with the minimum
of the potential at $\phi=v=\mu^2/\lambda$. There is no dependence 
of the potential on the angles $\alpha$, $\beta$ and $\gamma$.

The value of the potential at the minimum, where the Higgs field sits today,
is tightly constrained by cosmology. The potential acts like a cosmological constant term
in the gravitational field equations and leads to exponential expansion.
Whilst the current observational evidence is consistent with the universe
undergoing exponential expansion, the scale is tiny today compared to any energy 
scale relevant to the Higgs boson. 
Taking into account the zero potential at the minimum, the Higgs potential 
can be expressed succinctly as
\begin{equation}
V(\phi)=\frac14\lambda\left(\phi^2-v^2\right)^2.\label{higgspot}
\end{equation}
In quantum theory, the Higgs field becomes a quantum operator with the vacuum expectation
value set by the minimum of the potential, $\langle\phi\rangle=v$. The Higgs boson corresponds to an
excitation of the vacuum state with mass
\begin{equation}
m_H=\sqrt{V''(v)}=\lambda^{1/2}v,
\end{equation}
which the Large Hadron Collider has determined to be $126\,{\rm GeV}$.

The essential feature of the Higgs mechanism is what happens to the fields $\alpha$, $\beta$ and $\gamma$.
Although the Higgs boson is a neutral particle, the unified Higgs field of the Weinberg-Salam model is charged.
It couples to a $U(1)$ vector potential ${\bf B}$ with strength $g'$ and a triplet of $SU(2)$ vector potentials 
${\bf W}_1$, ${\bf W}_2$, ${\bf W}_3$, with strength $g$. The fields $\alpha$, $\beta$ and $\gamma$ are 
absorbed into a field redefinition of these vector potentials, ${\bf W}_i\to{\bf W}^\prime_i$.
This results in three massive vector fields $W_{\pm}$ and $Z$, with masses
 \begin{equation}
m_W=m_Z\cos\theta_W={gv\over 2}=80.4\,{\rm GeV},
\end{equation}
where $\tan\theta_W=g'/g$. One final combination
of the vector fields remains massless, and defines the electromagnetic potential
${\bf A}$,
\begin{equation}
{\bf A}={\bf B}\cos\theta_W+{\bf W}_3'\sin\theta_W,\label{emf}
\end{equation}
where ${\bf W}_3'$ is the re-defined ${\bf W}_3$. 
The elementary electromagnetic charge is related to the $SU(2)$ coupling by $e=g\sin\theta_W$.

All of the above is subject to the addition of quantum corrections to the potential,
the couplings and the relations between the couplings and the 
masses \cite{0954-3899-37-7A-075021}. As a result, the effective values of the
coupling constants become `running' coupling constants which depend on an energy 
scale. The gauge couplings $g$ and $g'$, for example, run at different rates
and appear to converge roughly to a common value at an energy scale around
$10^{15}{\rm GeV}$, suggesting that the symmetry becomes enlarged into a unified
symmetry group \cite{PhysRevLett.33.451}. 
The Higgs self coupling also runs, to low values at high energies, which is 
something we shall return to later.

\section{The inflationary Higgs}\label{sec3}

The power of modern cosmology is amply demonstrated in the way it removes
the possibility that the Standard Model Higgs field could be the field responsible for inflation. 
We shall see that the Higgs field is perfectly capable of producing inflation, but that resulting
large-scale structure of the universe is at odds with the high precision observations of the
large scale structure of the universe which are now possible.

To have inflation driven by the Higgs field, the Higgs potential energy density has to
come to dominate over everything else, and it has to hang around for long enough to  drive the
universe into exponential expansion. The theory of inflation gives a necessary
condition for this to happen in terms of a `slow-roll' parameter $\epsilon_V$ \cite{liddle2009},
\begin{equation}
\epsilon_V={M_p^2\over 2}\left({V'\over V}\right)^2<1.\label{epsilon}
\end{equation}
At large values of $\phi\gg v$, the Higgs potential (\ref{higgspot}) has a quadratic behaviour,
\begin{equation}
V(\phi)\approx \frac14\lambda\phi^4,\label{qpot}
\end{equation}
Thus $\epsilon_V\approx 8M_p^2/\phi^2$, and inflation could in principle take place for
$\phi>\sqrt{8}M_p$. Since the Higgs field contributes to particle masses, this runs
the risk of putting some of these masses beyond the Planck scale, introducing
quantum gravity effects. Any discussion of Higgs inflation is subject to some 
degree uncertainty for this reason.
   
The most remarkable feature of the inflationary scenario is the generation of
primordial density fluctuations from quantum fluctuations during inflation
\cite{1981ZhPmR..33..549M,guth82,hawking82-2,bardeen83,hawking83}.
The fluctuations can be separated into Fourier modes which track the expansion of the
universe, so that a mode with wave number $k$ grows in wavelength with the
growth of the universe, from around $10^{-28}\,{\rm m}$ during the inflationary
era to present-day cosmological scales measured in megaparsecs (Mpc). The evolution of 
the amplitude of these fluctuations into the temperature fluctuations seen in the Cosmic Microwave 
Background involves relatively low energy physics and is believed to  be well-understood \cite{liddle2009}. 
It is therefore possible to reconstruct features of the primordial fluctuations from the CMB,
and other large-scale structure observations.

Inflationary fluctuations which influence the CMB come in two types:
scalar and tensor. Scalar modes describe the fluctuations in the energy
density and the tensor modes are gravitational waves. Both types of
fluctuation have an effect on the local intensity and polarisation of the
CMB. The primordial power spectra of these fluctuations are conventionally
parameterised by
\begin{align}
{\cal P}_s&=A_s \left({k\over k_*}\right)^{n_s-1},\label{ps}\\
{\cal P}_t&=A_t \left({k\over k_*}\right)^{n_t-1},\label{pt}
\end{align}
where the spectral indices $n_s$ and $n_t$ give the leading order dependence on the wave number $k$.
A typical value used for the pivot scale is $k_*=0.05\,{\rm Mpc}^{-1}$. (The wavelength of the mode $k_*$
is presently around $126\,{\rm Mpc}$. To give some idea of scale, a cube of the corresponding size would 
contain around 10,000 galaxies.)
Recent data from the Planck satellite \cite{Ade:2015lrj,Ade:2015tva}, supported by a whole range of 
other observations, suggest that the scalar fluctuations predominate, with $A_s\approx 2.25\times 10^{-9}$.
The tensor contribution, tracked by the tensor to scalar ratio $r={\cal P}_r/{\cal P}_s$, is less than
10\% of the total. 
  
The theory of inflationary fluctuations gives us the scalar fluctuation amplitude in terms of the
potential and the slows-roll parameter $\epsilon_V$ as
\begin{equation}
A_s={1\over 24\pi^2}{V_*\over M_p^4\,\epsilon_{V*}}.\label{as}
\end{equation}
The wavelength of a fluctuation of wave-number $k$ actually increases with the expansion 
of the universe. The star on the potential $V_*$ indicates that the potential is evaluated at a time 
$t_*$ when the wavelength of the expanding fluctuation equals the natural length scale, 
or Hubble radius, of the inflationary universe.
The logarithmic growth of the wavelength between $t_*$ and the end of inflation
is an important number, called the `number of e-folds', $N_*$. This is related to the
slow-roll parameter by $N_*\approx 1/\epsilon_{V*}\gg1$, allowing us to 
express the amplitude (\ref{as}), using (\ref{epsilon}) and (\ref{qpot}), in terms of $N_*$,
\begin{equation}
A_s={2\over 3\pi^2}\lambda N_*^3.\label{asn}
\end{equation}
The number of e-folds is set by the requirement that the density fluctuation has to grow in wavelength
from the Hubble radius $10^{-28}\,{\rm m}$ at the time $t_*$ during inflation, to $0.05\,{\rm Mpc}$ today. 
This requires inflationary models to have $N_*$ between 50 and 60, depending on how the universe evolves after
the end of the inflationary era.
We know $A_s$ from the size of the temperature fluctuations in the CMB, and combining this with 
$N_*>50$ gives an incredibly tight limit on the coupling $\lambda<0.3\times 10^{-12}$. 
We will see in Sect \ref{unstab} that the Higgs coupling becomes small at high values of the Higgs field due to
quantum effects, but even so such a small value seems implausible. 

The final, and conclusive, blow to this simplest type of Higgs inflation comes from the ratio
of tensor to scalar fluctuations. Inflationary theory gives the tensor amplitude
\begin{equation}
A_t={2\over 3\pi^2}{V_*\over M_p^4}.\label{at}
\end{equation}
The ratio of (\ref{at}) to (\ref{as}) is therefore
\begin{equation}
r_*=16\epsilon_{V*}={16\over N_*}>0.27,
\end{equation}
for $N_*<60$. This conflicts with the limit $r<0.1$ obtained from observations of the CMB and
the large-scale galaxy distribution. 

Besides simple Higgs inflation, this argument shows that any type of inflation based the relativistic 
wave equation and a quartic potential is in conflict with the observations.
The argument has to be reconsidered if there are modifications to the derivative terms in the theory,
or coupling to other scalar fields with a different potential. This happens in the variations of 
Higgs inflation described below.
Nevertheless, ruling out simple inflationary models with quartic potentials is a 
remarkable achievement of observational cosmology. 

\subsection{Type I Higgs inflation}

The major omission from the Standard Model of particle physics is gravity.
We might expect the elementary particles of the Standard Model to satisfy the basic principles of relativity,
and couple to the geometry of spacetime in a way which respects the
equivalence principle, but this leaves open a wide range of possible Higgs models. 
Two types of Higgs inflation take advantage of this ambiguity in the way in which the Higgs couples to gravity. 

The oldest type of Higgs inflation includes a coupling between the Higgs field and the
Ricci scalar $R$ of the spacetime geometry 
\cite{Futamase:1987ua,Salopek:1988qh,Fakir:1990eg,Kaiser:1994vs}. 
This interaction is given by a 
Lagrangian density ${\cal L}_I$ with a parameter $\xi$,
\begin{equation}
{\cal L}_I=\xi R\,{\bf H}^\dagger{\bf H}.
\end{equation}
There are no compelling reasons for omitting a curvature-coupling term like this from the 
scalar field Lagrangian density (apart from technical difficulty). 

The curvature-coupling term couples the metric and the scalar field fluctuations. After these
have been separated, the scalar fluctuation amplitude given previously by (\ref{as})
now becomes \cite{Bezrukov:2013fka},
\begin{equation}
A_s={2\over 3\pi^2}\lambda{N_*^3\over (1+8\xi N_*)(1+6\xi)}.
\end{equation}
The limits on $A_s$ set by the CMB can now be realised provided
that $\xi\sim 10^5$, maybe rather large but not as unreasonable as the
previous restrictions on $\lambda$.\footnote{These calculations usually proceed by
using a fictitious metric called the Einstein frame \cite{Bezrukov:2013fca}. 
The results here have been calculated using the physical metric of the Standard Model, 
or the `Jordan frame'.}

\begin{figure}
\begin{center}
\resizebox*{10cm}{!}{\includegraphics{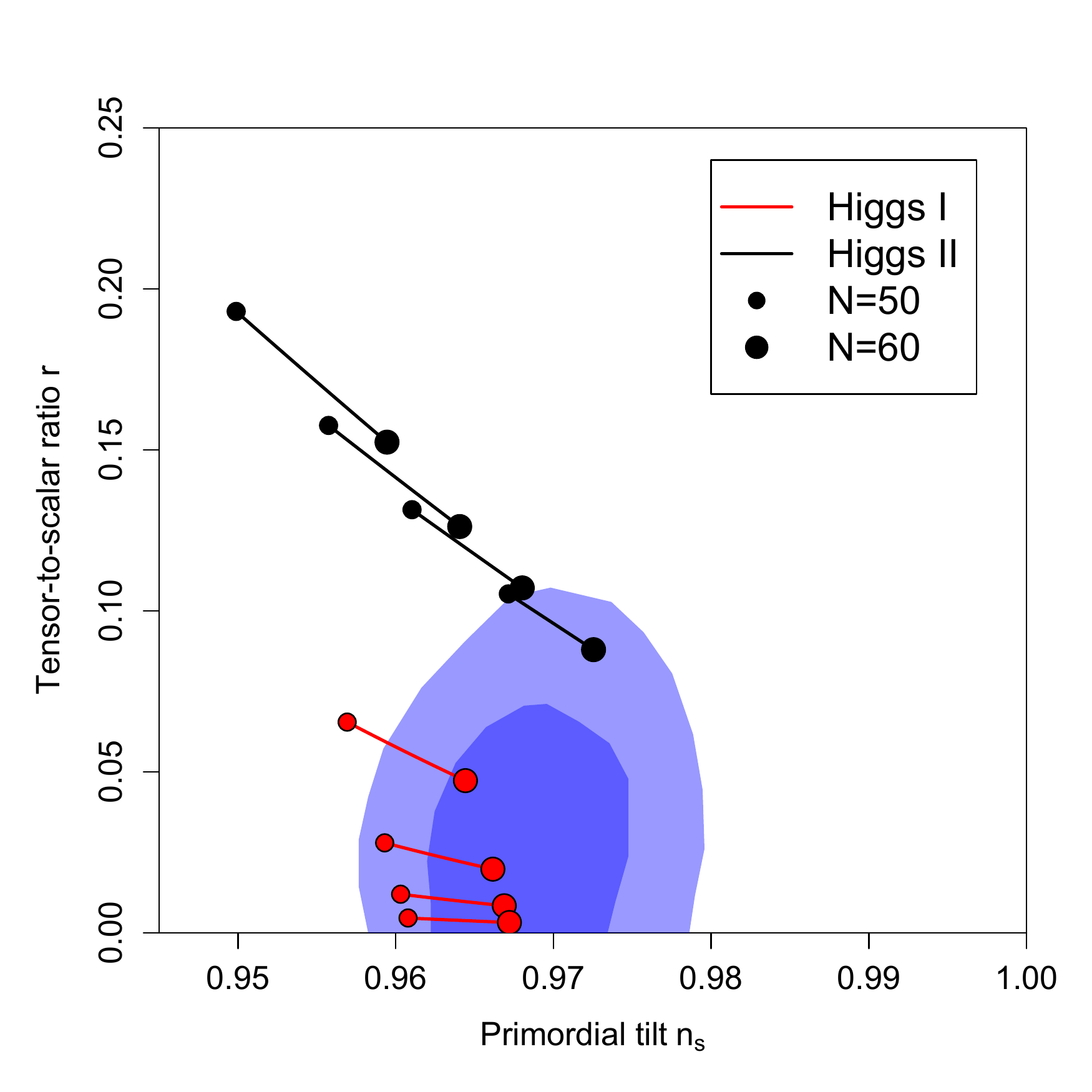}}
\caption{Fluctuations in the CMB can be used to infer the primordial amplitude ratio of tensor to 
scalar modes $r$ and the spectral index of the scalar fluctuations $n_s$. 
The predictions for two types of Higgs inflation are 
shown, with a selection of $\xi$ and $w$ parameters (see text). The solid areas are the $67\%$ and
$95\%$ confidence regions copied from the Planck publication \cite{Ade:2015lrj}, which
combines CMB and other large scale structure data to obtain limits on $r$ and $n_s$.} \label{rns}
\end{center}
\end{figure}

The tensor-scalar ratio $r$ and the spectral index of the scalar fluctuations $n_s$
can be calculated for this model and, if we set aside the question of quantum gravity 
corrections to the Higgs potential,
\begin{align}
r_*&={16\over N_*}{1+6\xi\over 1+8\xi N_*}\\
n_s&=1-{3\over N_*}{3+16\xi N_*\over 3+24\xi N_*}.
\end{align}
These are plotted in figure \ref{rns}, which shows how $r$ decreases as $\xi$ ranges from 
$0.01$ to $100$. (The plots of $\xi>100$ are almost identical to the $\xi=100$ case.) 
The original Higgs model with $\xi=0$ gives values for $r$ and $n_s$ which
are outside of the range shown by this plot. The agreement with the observational limits for
the wide range of values for $\xi$ in this plot is impressive.

This model is not without its problems, however. The large value of $\xi$ increases
the effect of quantum gravity corrections to the Higgs potential at values of the Higgs
field encountered during inflation
\cite{Burgess:2009ea,Barbon:2009ya,Burgess:2010zq,Hertzberg:2010dc,Atkins:2010yg}.
It seems, though this is still controversial, that these 
corrections may affect the values of $r$ and $n_s$
\cite{Bezrukov:2013fka,Moss:2014nya}. It also has to be mentioned that there are many other 
models of inflation which
produce agreement with the observed limits on $r$ and $n_s$  \cite{Ade:2015lrj}. 
The special feature of Higgs inflation is that it is based on a field which we already know exists.

\subsection{Type II Higgs inflation}

The interactions between the Higgs field and gravity have one further restriction, and this
is a limit on the number of derivative terms appearing in the Lagrangian density. Having too many
derivatives runs the risk of producing an unstable theory with no ground state, because some of the kinetic
terms in the energy can become negative. The second type of 
Higgs inflation is the next simplest which avoids this problem. It includes a coupling between the 
Higgs field and the Einstein tensor $G^{\mu\nu}$ of the spacetime geometry
\cite{Germani:2010gm,Germani:2010hd,Germani:2011ua}.
The interaction is given by a Lagrangian
density with a parameter $w$,
\begin{equation}
{\cal L}_I=-w^2 G^{\mu\nu}(D_\mu{\bf H})^\dagger (D_\nu{\bf H}),
\end{equation}
where $D_\mu{\bf H}$ is the $SU(2)\times U(1)$-symmetric derivative of the Higgs field with respect to 
the spacetime coordinates and $\mu$ runs from $0\dots 4$.

This new type of curvature coupling again changes the amplitude of density fluctuations, 
replacing the old result (\ref{as}) by 
\begin{equation}
A_s={1\over 24\pi^2}{V_*\over M_p^4}{(1+Q_*)\over\epsilon_{V*}},\label{asII}
\end{equation}
where $Q_*=w^2 V_*/M_p^2$ reflects the change to the Higgs derivative terms. 
The relationship between the the slow-roll parameter and the number of e-folds
is also modified to $\epsilon_{V*}\approx(1+Q_*/3)/N_*$. The tensor-scalar ratio and 
the spectral index of the scalar perturbations $n_s$ for this model become,
\begin{align}
r_*&={16\over N_*}{1+Q_*/3\over 1+Q_*},\\
n_s&=1-{3\over N_*}{(1+5Q_*/3)(1+Q_*/3)\over(1+Q_*)^2},
\end{align}
where $Q_*\propto (wN_*)^{2/3}$.
These are plotted in figure \ref{rns}, with values of $w$ ranging from $0.1\,{M_p}^{-1}$
to $100\,{M_p}^{-1}$ (from top to bottom). The agreement with the observational limits is 
not as good as before, there is still consistency for large $w$.

\section{The electroweak Higgs}\label{sect3}

The Higgs field plays the central role in the electroweak phase transition at a temperature
around $163\,{\rm GeV}$ and a time around one nanosecond after the big bang. 
Just prior to this transition, particle interaction timescales 
are far shorter than the timescale of the evolving universe, and the radiation which fills the
universe is, in effect, in perfect thermal equilibrium. What happens next depends on the
details of the Higgs potential. 

\begin{figure}
\begin{center}
\resizebox*{10cm}{!}{\includegraphics{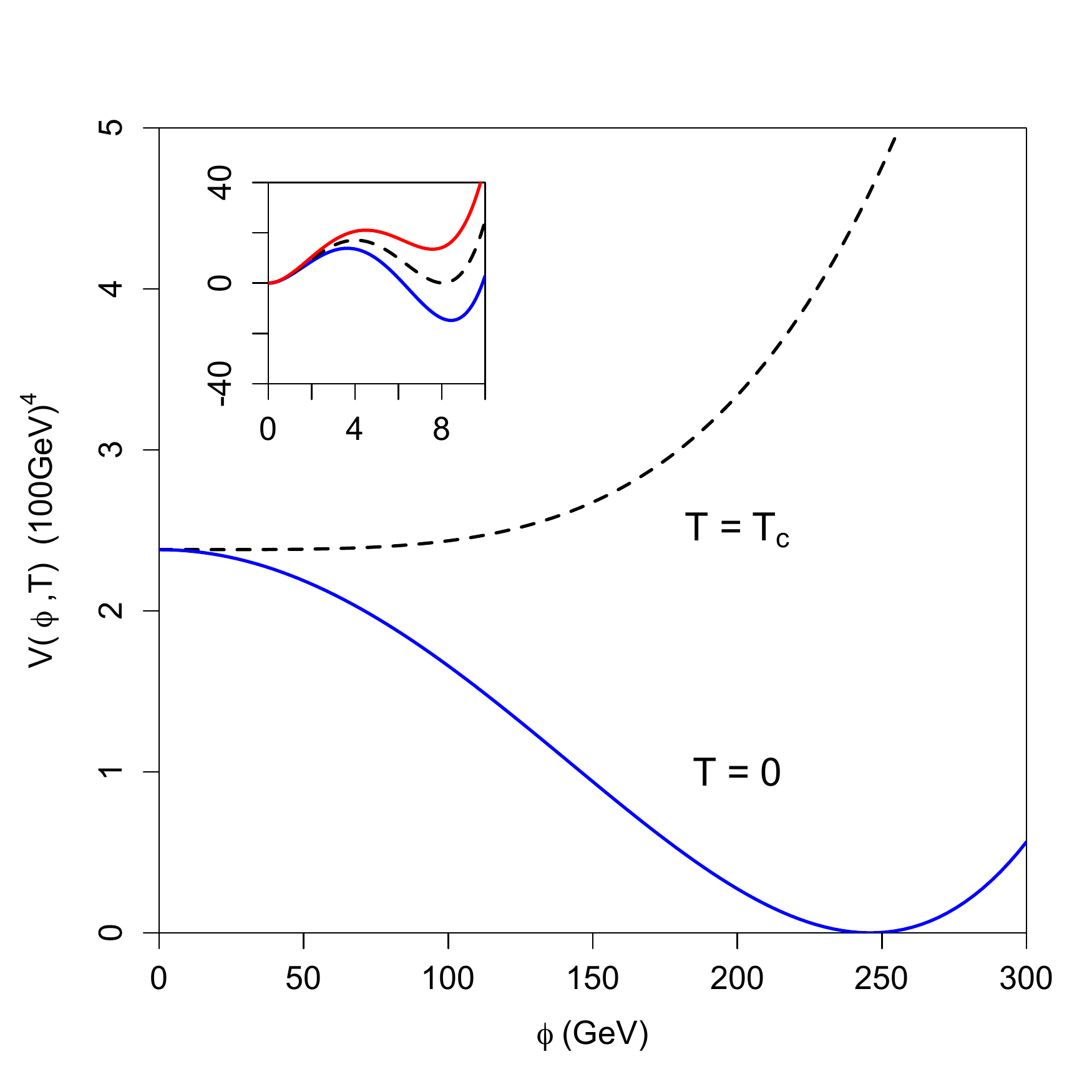}}
\caption{The temperature-corrected Higgs potential implies symmetry restoration at high
temperatures. The inset shows the region near the origin, for three temperatures close to
$T_c\approx 163\,{\rm GeV}$, and a magnified vertical scale in $({\rm GeV})^4$.} 
\label{ewp}
\end{center}
\end{figure}

Thermal particle interactions with the Higgs field introduce temperature dependent terms
into the Higgs potential. The leading order terms for the Standard Model are
(see e.g. \cite{Katz:2014bha}),
\begin{equation}
V(\phi,T)\approx V_0(T)+\frac12m^2(T)\phi^2-ET\phi^3+\frac14\lambda(T)\phi^4,
\end{equation}
where $m(T)$ and $\lambda(T)$ are an effective thermal mass and coupling of the Higgs field, and
\begin{equation}
E={1\over 6\pi v^3}\left(2m_W^3+m_Z^3\right)\approx0.0064.\label{eee}
\end{equation}
This potential is shown in figure \ref{ewp}. At a critical temperature $T_c$ the potential has
two minima at $\phi=0$ and $\phi=\phi_c$ with equal potential $V(0)=V(\phi_c)$. 
Above this critical temperature, the potential has a global minimum at the symmetric point $\phi=0$. 
The SU(2) vector bosons, for example, have equal thermal masses. 
We can think of this as the symmetric phase. At low temperatures, the Higgs field is close to 
its vacuum expectation value $v$, and the particle spectrum is characteristic of the broken 
symmetry phase.

The potential has a very small barrier near the critical temperature. However, 
the thermal fluctuations in the Higgs field
$\delta\phi\sim T_c$ swamp the barrier, which has width $\phi_c=2ET_c/\lambda\sim10^{-2}T_c$
\cite{Borrill:1994nk}. 
In this case, as the temperature drops, the thermal average of the Higgs field 
evolves continuously from the symmetric phase into the the broken symmetry phase, 
maintaining local thermal equilibrium.

The Higgs field also has the three additional components $\alpha$. $\beta$ and
$\gamma$, which eventually become absorbed by the vector potential fields. The initial values of these fields
are correlated on small scales but essentially random on large scales and they will evolve to 
different values in widely separated regions of the 
universe which are out of causal contact with one another.
The electromagnetic field in different parts of the CMB (see Eq. (\ref{emf})) is initially made up from different 
combinations of the 
SU(2) and Higgs fields. The large scale fluctuations in $\alpha$, $\beta$ and $\gamma$ create large scale 
electric and magnetic field fluctuations, but the field strength is 
given by a gradient of the Higgs fluctuations and because of the large length scale of the fluctuations
the field is too small to be observed.
 
The continuous electroweak phase transition of the Standard Model seems sadly devoid 
of any observable consequences. A first order electroweak phase transition, on the other hand, could
have significant observational effects such as:
\begin{enumerate}
\item The generation of baryon asymmetry. Expanding bubbles of broken symmetry phase combine with
CP symmetry violation to produce a net excess of baryons over anti-baryons  
\cite{Shaposhnikov:1987tw,Farrar:1993sp,Kuzmin198536}.
\item The generation of gravitational waves. Expanding bubbles of broken symmetry phase act as sources of 
gravitational waves \cite{Grojean:2006bp}.
\item The generation of magnetic fields. Colliding bubbles of broken symmetry phase produce a
turbulent dynamo \cite{Baym:1995fk}.
\end{enumerate}
Adding extra bosons to the model increases the parameter $E$ in Eq. (\ref{eee}), possibly raising the height of the 
potential above the magnitude of thermal fluctuations and making the phase transition first order. These extra bosons can 
come from supersymmetric extensions of the Standard Model \cite{Carena:2004ha} and/or additional 
Higgs fields \cite{Turok1992729}. Therefore remnants from the electroweak phase transition might possibly 
provide evidence of new physics at the TeV scale.

\section{The unstable Higgs}\label{unstab}

Our final look at Higgs cosmology considers whether the Higgs field might undergo a phase transition at
some time in the future. At the present time the Higgs field would be in a supercooled state,
analogous to a bottle of pure water cooled a few degrees below zero Celsius in the freezer.
If the water is gently removed from the freezer it sometimes continues in a 
supercooled liquid state. Any violent disturbance to the bottle causes a rapid transition 
from water to ice.

The question of Higgs metastability arises even in the Standard Model \cite{Sher:1988mj}. Quantum effects
modify the Higgs potential, producing an effective potential $V_{\rm eff}$.
At large field values, the potential can be expressed in terms of an effective quartic coupling 
$\lambda_{\rm eff}(\phi)$,
\begin{equation}
V_{\rm eff}(\phi)\approx\frac14\lambda_{\rm eff}(\phi)\phi^4.\label{effp}
\end{equation}
If $\lambda_{\rm eff}$ becomes negative, then the state with $\langle\phi\rangle=v$ is no longer
the lowest energy state. This state is a `false vacuum' state, which can be at best metastable.

The stability of the Higgs potential is sensitive to 
the Standard Model parameters, especially the Higgs mass and the top quark mass 
(which being the most massive fermion is the one with the largest coupling to the Higgs field). 
In first-order perturbation theory, 
each bosonic mode with frequency $\omega$ contributes 
a vacuum energy $\hbar\omega/2$ to the potential, and each fermionic mode contributes
$-\hbar\omega/2$, due to the fermionic particle statistics. This effect of this sign difference
persists to higher orders in perturbation theory, so that a large top quark mass has a destabilising
effect on the Higgs potential. 

\begin{figure}
\begin{center}
\begin{minipage}{140mm}
\subfigure[The running coupling constant.]{
\resizebox*{7cm}{!}{\includegraphics{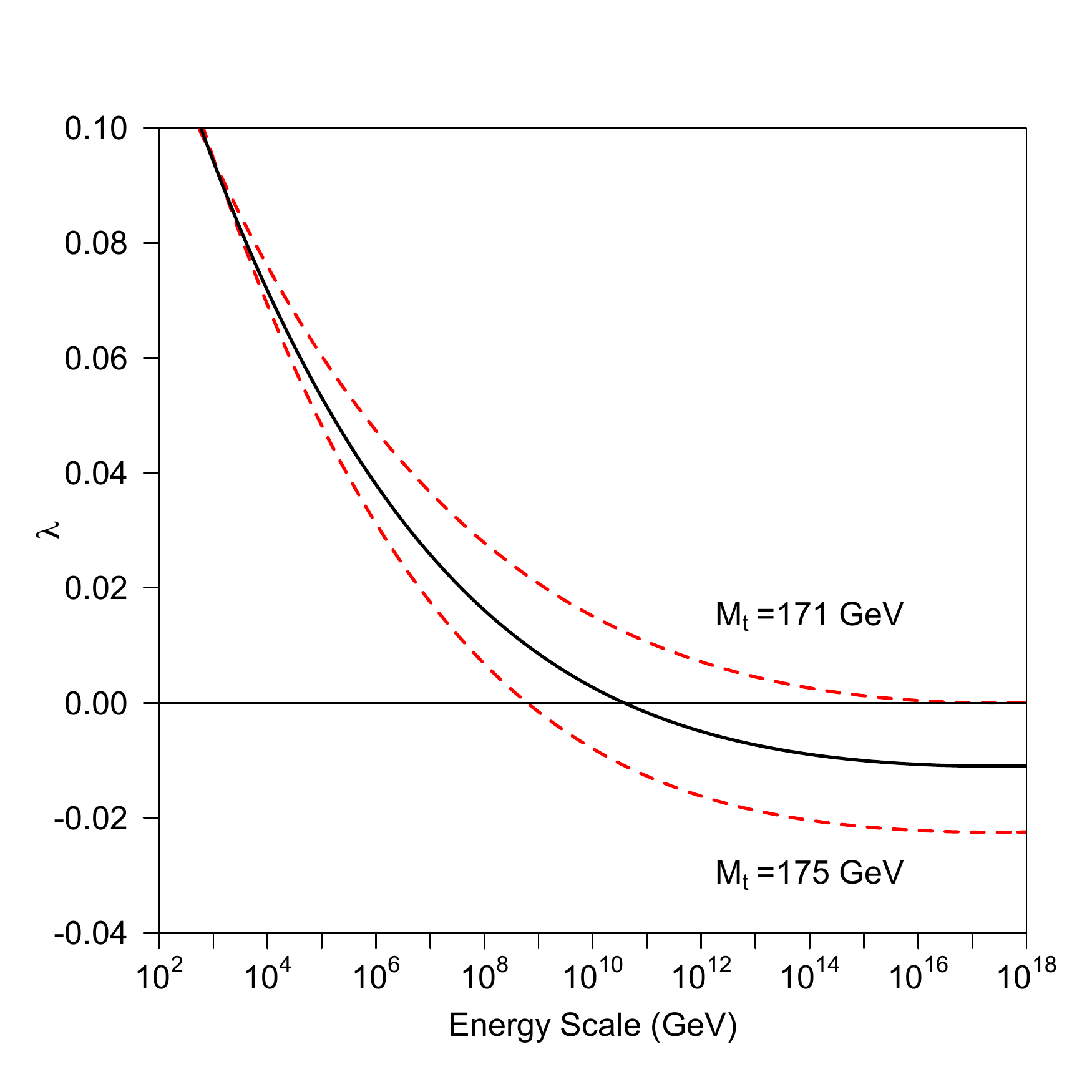}}}\hspace{6pt}
\subfigure[The unstable Higgs effective potential.]{
\resizebox*{7cm}{!}{\includegraphics{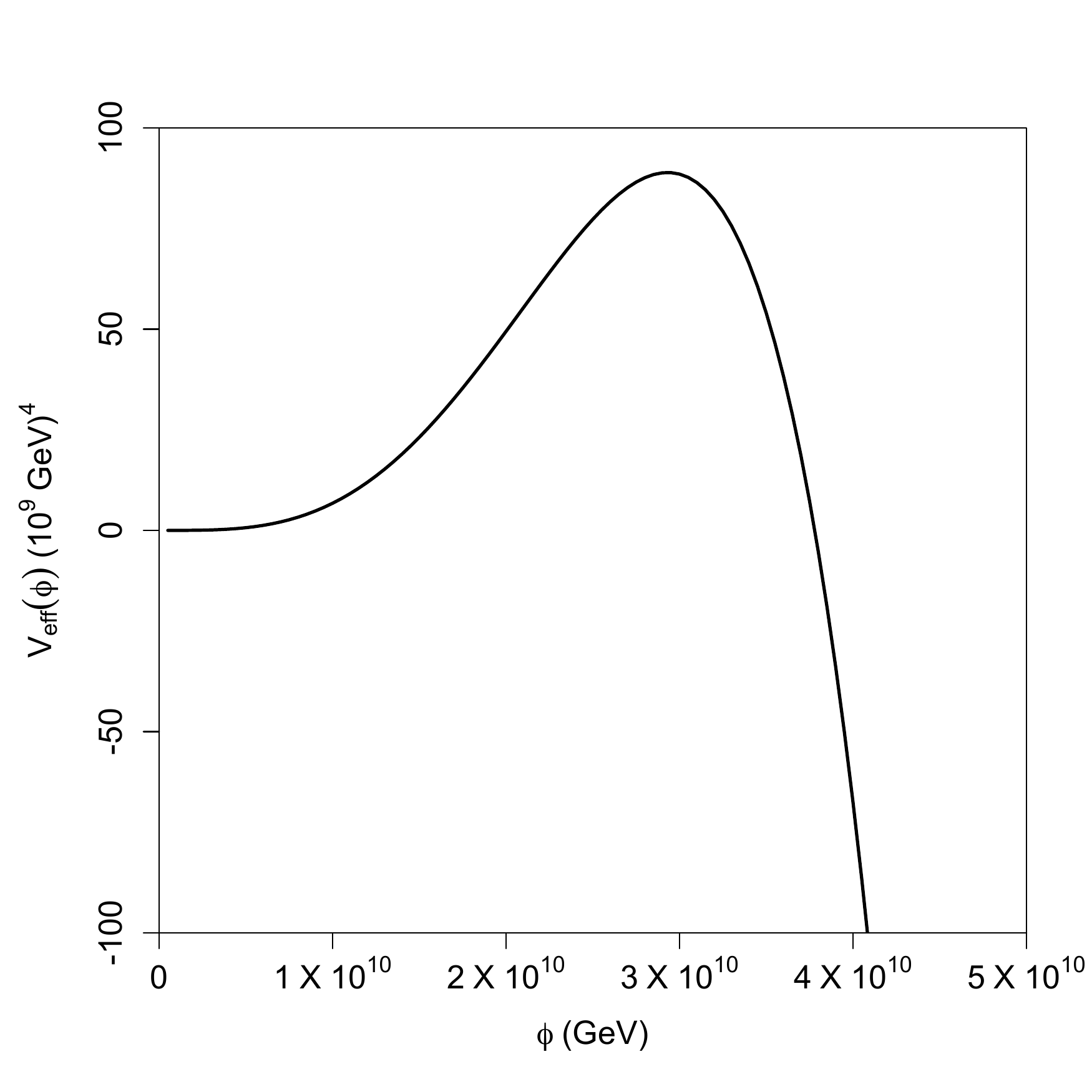}}}
\caption{(a) Shows a numerical fit to the running coupling constant for the Higgs field of the
standard model with $M_H=125$GeV. The running coupling appears to become negative at high energies, 
resulting in a metastable Higgs potential (b). 
The coupling is shown for a range of the top quark mass $M_t$.} \label{lambda}
\end{minipage}
\end{center}
\end{figure}

Figure \ref{lambda} shows the running 
coupling $\lambda(\mu)$ in second-order perturbation theory \cite{Degrassi:2012ry}.
The running coupling $\lambda(\mu)$ is closely related to the effective coupling,
very roughly $\lambda_{\rm eff}(\phi) \sim \lambda(\phi)$ with a quantifiable correction.
The running coupling becomes small at large values of the Higgs field, and runs negative for 
a Higgs mass $m_H=125\,{\rm GeV}$ and top quark masses above $171\,{\rm GeV}$.
Measurements of the top quark mass, $173.34\pm0.76\,{\rm GeV}$ \cite{ATLAS:2014wva},
place it in firmly the unstable regime. The effective potential (\ref{effp}) for this range of parameters is 
also shown in Figure \ref{lambda}.

\begin{figure}
\begin{center}
\resizebox*{10cm}{!}{\includegraphics{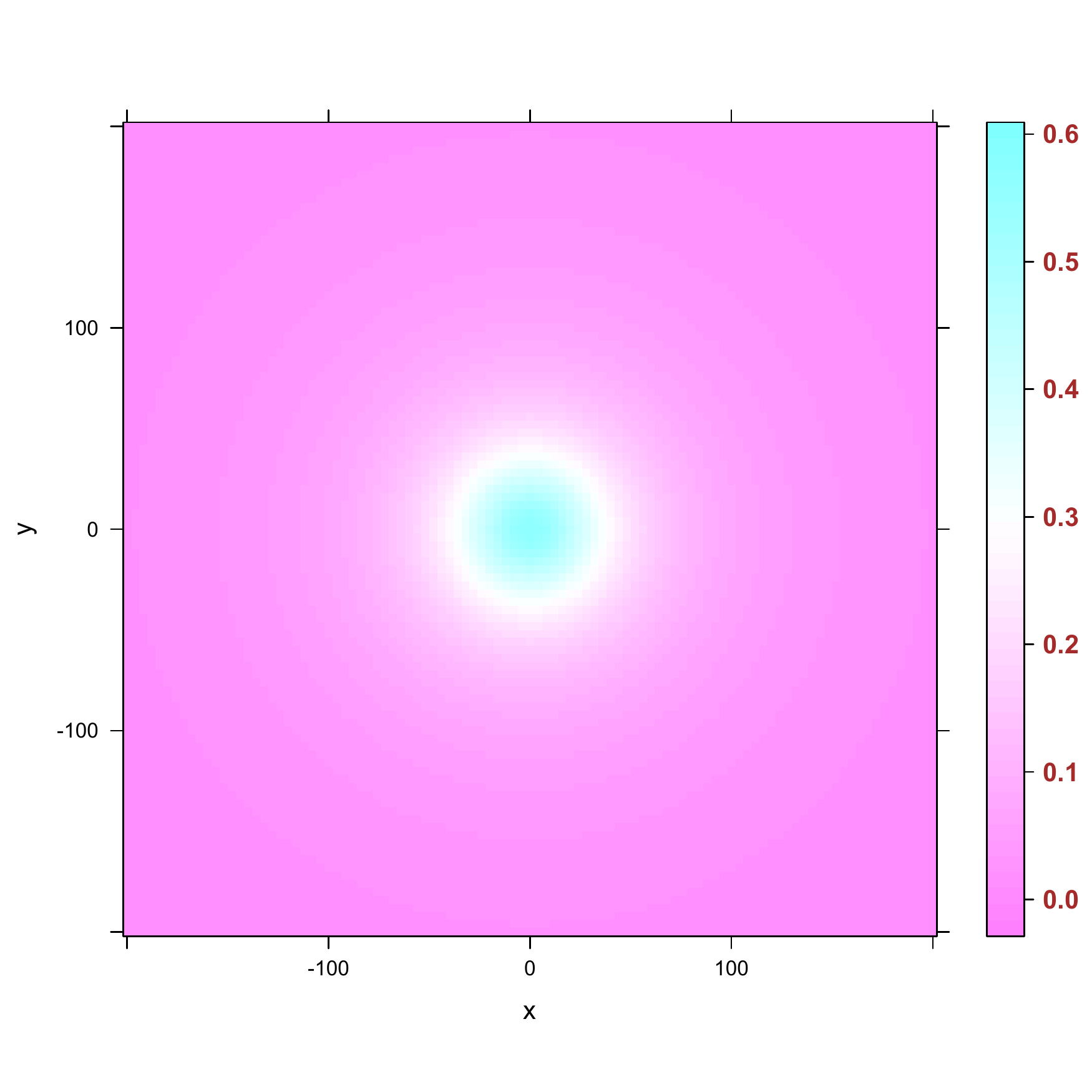}}
\caption{The spatial distribution of  the Higgs field in a vacuum decay bubble. Two dimensions are shown, 
the bounce solution has the same profile in all four dimensions (three space and one imaginary
time). The central region of the bubble has 
large values of $\phi$ stretching well beyond the potential barrier from the false vacuum (pink) 
into a new Higgs phase (blue).  All measurements are in reduced 
Planck units. The effective coupling here is modelled by 
$\lambda_{\rm eff}=-0.02+10^{-4}(\ln\phi/M_p)^2$ and the instanton action is calculated numerally to be
$B\approx1360$. } 
\label{ew}
\end{center}
\end{figure}

The unstable false vacuum can decay by quantum tunnelling through the potential barrier, producing
bubbles filled with Higgs field at a potential lower than the energy of the false vacuum. The rate of
bubble nucleation is determined by a procedure adapted from the theory of supercooled first order
phase transitions. Recall that, for phase transitions, the probability of bubble nucleation
is proportional to $e^{-\Delta S}$, where $\Delta S$ is the change in entropy due to the bubble. 
For vacuum decay, one first solves the field equations with an imaginary time coordinate $\tau=it$, 
to obtain a solution which Coleman called the `bounce'  \cite{coleman1977,callan1977,Coleman1980}. 
The bubble nucleation rate is given by
\begin{equation}
\Gamma_D= A e^{-B},\qquad A\approx {B^2\over 4\pi^2 r_b^4},\label{cab}
\end{equation}
where $B$, the action of the bounce solution, plays the same role as the entropy $\Delta S$ in
a phase transition. The pre-factor $A$ is related to the bubble radius $r_b$. The Higgs field profile 
of a bubble after it nucleates matches the $\tau=0$ slice through the bounce 
solution. After nucleation, these bubbles are unstable and they grow, eventually expanding at the
speed of light. The interior of the bubble sinks towards the true Higgs vacuum,
and the spacetime geometry inside becomes wildly distorted, whilst outside the expanding bubble
eats into the false vacuum state.

An example of a bubble solution to the coupled Higgs and gravitational field equations is shown in 
Figure \ref{ew}. A single bubble of this type would eventually switch the vacuum state of the
observable universe. Luckily, the nucleation rate for this type bubble is entirely negligible. Using formula
(\ref{cab}), the pre-factor for the bounce solution shown in Figure \ref{ew} is $A\sim 10^{69}\,{\rm GeV}^{4}$, 
but the action $B\approx 1320$ and the exponential term in the nucleation rate dominates.

On the whole, vacuum decay rates tend to be very strongly exponentially suppressed. However,
in nature most phase transitions take place due to some type of nucleation seed, an impurity
or an imperfection an a containment vessel. There has been some discussion recently as to
whether a tiny black hole could act as a seed and trigger false vacuum decay \cite{Burda:2015isa}.
Small black holes may be produced in the early universe \cite{Khlopov:2008qy}, or in a high energy accelerator
\cite{Dimopoulos:2001hw,Gregory:2008rf,Kanti:2014vsa}.

\begin{figure}
\begin{center}
\resizebox*{10cm}{!}{\includegraphics{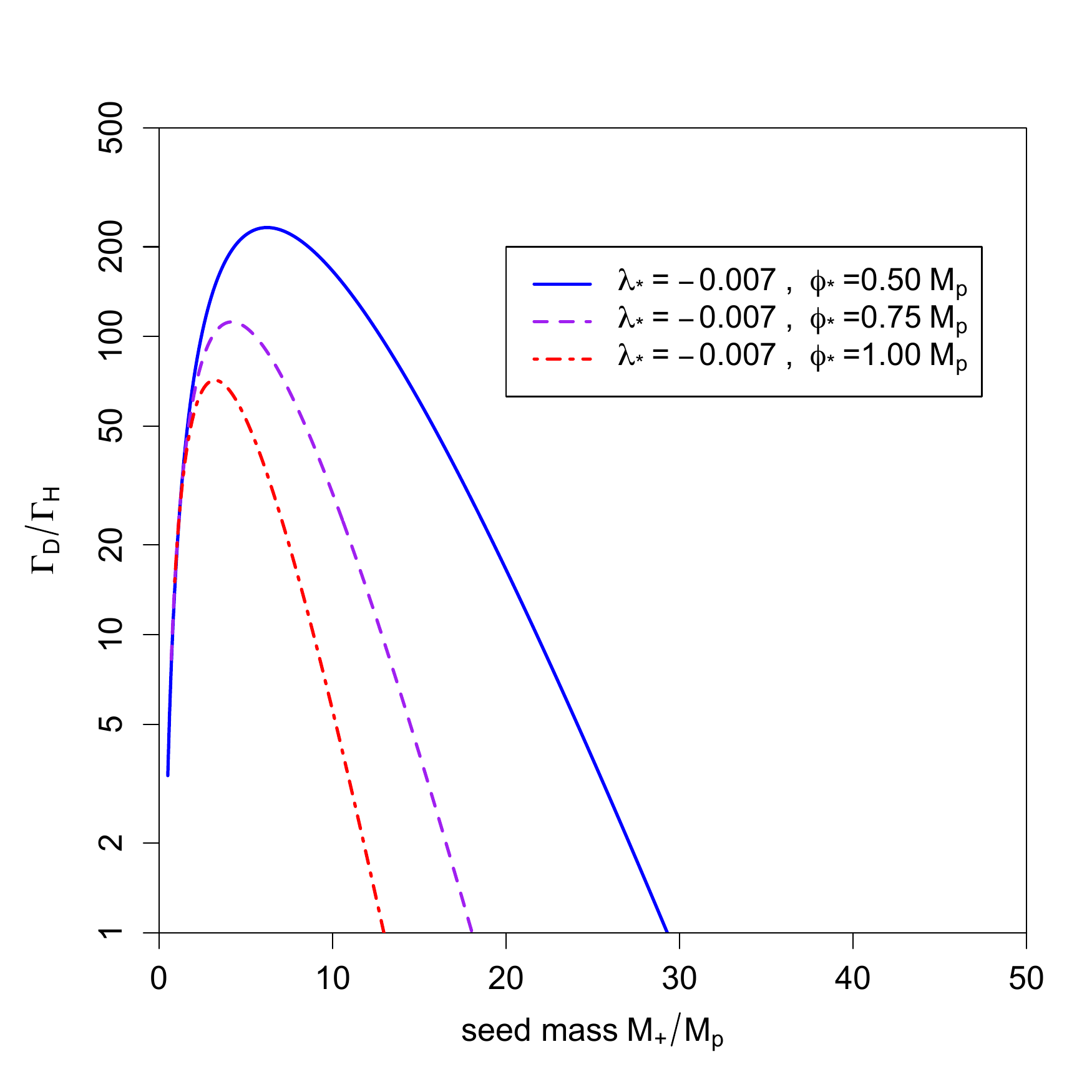}}
\caption{The ratio of the rate of vacuum decay rate caused by a black hole $\Gamma_D$
to the Hawking evaporation rate of the black hole $\Gamma_H$ for different black hole
masses $M_+$. The effective coupling has been is 
modeled by $\lambda_{\rm eff}=\lambda_*+10^{-4}(\ln\phi/\phi_*)^2$, and the Hawking evaporation rate 
here is $\Gamma_H\approx 3.6\times 10^{-4}M_p^4/M_+^3$. For these model parameters, a
primordial black hole seeds vacuum decay after it reaches a mass $10-30\,{\rm M_p}$.} 
\label{bh}
\end{center}
\end{figure}

One obstacle to black holes as nucleation seeds is that black holes evaporate by the Hawking process,
and small black holes evaporate very rapidly. Therefore it is important to compare the
seeded nucleation rate to the rate of Hawking evaporation. This has been done in Figure
\ref{bh}, where we see that the vacuum decay rate beats the Hawking evaporation rate for small holes with
masses 10-100 times the reduced Planck mass.

If we start with a primordial black hole of mass around $10^{12}\,{\rm kg}$ produced in the early universe, it
takes a few ${\rm Gyr}$ before it decays to the relevant mass range. At this point, it can nucleate
the vacuum decay bubble and initiate the phase transition. Clearly, no such phase transition
has taken place, therefore the metastable Higgs vacuum is inconsistent with the existence of
even a single primordial black in our observable universe. This is puzzling, because the standard
model Higgs potential produces an unstable Higgs potential with the presently measured values of the
top quark and Higgs boson mass. The resolution may be that there are stabilising contributions to the Higgs
potential from physics beyond the standard model.

There is another way in that tiny black holes can occur, which requires the existence of
extra dimensions. In the relevant theoretical framework, the Standard Model particles lie on a surface, 
or `brane', in higher dimensions. The higher-dimensional Planck mass has a similar scale to the Standard Model,
but becomes exponentially enlarged on our brane. Black holes can
then be produced in particle collisions with energy far below the three-dimensional Planck mass, 
possibly even at the scale of the Large Hadron Collider.
This raises an alarming prospect of triggering vacuum decay at the LHC, but
fortunately, we have some re-assurance from the fact that cosmic ray collisions have occurred in 
nature at energies higher than those that can be reached at the LHC \cite{Hut1983}. Although 
`head on' cosmic ray proton collisions with energy in the TeV range or larger are incredibly rare
by particle collider standards, they are quite common over cosmological length and time scales 
relevant for cosmological vacuum decay. Cosmic ray collisions have not triggered vacuum decay,
and neither should the LHC. Once again, microscopic black holes are incompatible with unstable Higgs
potentials.

\section{Outlook}

Our short tour of Higgs cosmology has been largely confined to
the the particle content of the Standard Model plus gravity. We have not said anything
about the possibilities of new physics in the desert between $100\,{\rm GeV}$ and the scale
of Grand Unification, or indeed about Higgs fields associated with Grand Unification.
Some of the ideas mentioned here can be carried across directly to the GUT Higgs
field. Models of Higgs inflation work equally well with GUT Higgs fields, and the
flexibility in the Higgs potential allows a wider range of inflationary scenarios.

We can be confident, at the very least, that quantum gravity corrections will
have an effect both on the Higgs inflationary scenarios and the Higgs stability
question. Since we are working up from low energies towards the Planck scale,
it seem likely that perturbative quantum gravity is a good place to begin
\cite{Burgess:2003jk}.
This speculative end of Higgs physics is therefore providing an impetus
for new developments in quantum gravity. Finally, Higgs cosmology offers the
possibility that we shall see observational clues, gravity waves from
the electroweak phase transition or the detection of primordial black holes,
which point the way to physics beyond the standard model.

\section*{Acknowledgements} 

Section 5 contains some work done in collaboration with
Philipp Burda and Ruth Gregory. IGM is supported in part by STFC 
(Consolidated Grant ST/J000426/1).

\section*{Notes on contributor}

Ian Moss is Professor of Theoretical Cosmology at Newcastle
University, UK. He obtained his PhD from Cambridge under
the supervision of Professor Stephen Hawking in 1982.
The start of his career coincided with the birth of cosmological inflation theory,
and began with colliding bubbles and Hawking-Moss instantons.
His personal inflationary trajectory continues up to the present day, with some
diversions into other aspects of relativity, black holes and quantum gravity 
along the way.

\bibliographystyle{tCPH}
\bibliography{paper.bib}

\end{document}